# A wide band low profile linear cross polarizer for millimeter wave applications

**SHIVA HAJITABARMARZNAKI, MEHDI AHMADI-BOROUJENI, RANA NAZIFI, SEYEDEHZAHRA SHOJAEIAN, IMAN MIRZAEI, MEHDI FARDMANESH**

*Department of Electrical Engineering, Sharif University of Technology, Tehran, Iran.*
*\*ahmadi@sharif.*

**Abstract:** In this article, two wideband high-efficiency metasurface structures as a linear-cross polarizer in transmission mode based on asymmetric split-ring resonators (ASRR) are proposed. A unit cell of the first structure consists of double polarization-sensitive S-shape resonators (SRs) on the bottom side of the dielectric layer. The superiority of this proposed structure is its filtering property and high conversion efficiency at its resonant frequency, 100 GHz, which its cross-polarized transmission coefficient reaches 0.87. The simulation results show that the polarization conversion ratio (PCR) exceeds 85% in the frequency range from 86 to 139 GHz. Additionally, the second structure consisted of a wire-grid array as a bottom layer. This proposed structure has extremely low co-polarized transmission amplitude which represents exceedingly a good polarization transformer in cross-direction. Furthermore, this structure has a polarization conversion ration more than 95% in frequency from 50 to 200GHz, having wider relative bandwidth compared to the first proposed polarizer. The performance of the both polarizer have very weak angular dependency under the oblique incident of *x*-polarized EM wave with $\theta$ from $0°$ to $50°$. A prototype of these proposed polarizers has been fabricated and a good agreement between experimental and simulation results has been observed. These proposed structures can find application in mm-wave and THz sensing and imaging systems which benefit from polarimetric measurements.

## 1. Introduction

In recent years, the ability to controlling the polarization state of EM waves [1] has intrigued great attention due to its considerable application in imaging systems [2], imaging polarimetry [3], changing the polarization of antennas [4], and THz wireless communications [5]. To actualize these applications, designing low-profile polarization-converter with high-performance has been extremely investigated by researchers. In the past decade, polarizers were conventionally made of natural birefringent materials [6] or liquid crystals [7], which decompose EM wave into two orthogonal components with phase retardations. However, large loss, huge volume, and narrow bandwidth of such natural materials have limited their applications in integrated systems and THz applications.

More recently, artificial structures have been used for polarization transformation, which has exceptional capabilities in overcoming the physical limitation of these natural materials. In fact, Metasurfaces have been reported to provide a new pathway in artificially manipulating the polarization states of EM waves thanks to their outstanding features like low loss, compact size, and high efficiency. Planar metamaterials are composed of periodic and quasi-periodic planar arrays of sub-wavelength metallic or dielectric structures that can be fabricated by advanced technologies such as lithography and nanoprinting methods [8-9]. These promising materials have inevitable capabilities in controlling the amplitude, phase, and polarization state of EM waves in a wide range of frequency from microwave to optical regions. Numerous novel devices such as polarization converter [27-38], polarization beam splitters [10], vortex plates [11], lenses [12], phase modulators [13], and absorbers [14] show the potential applications of this planar metamaterial in miniaturized optical technologies.

As mentioned above, one of the most vital applications of metasurfaces are toward the realization of polarization manipulation of EM waves. Several kinds of multi-layer metasurface structures have been proposed as a high-efficiency polarization converter [15-22]. Some of these polarizers operate in reflection mode [16,19,22-24]. For instance, Kai-kai Xu et al. [19] proposed a reflective polarization converter based on double-split ring resonators (DSRR) in the THz band. This structure has a cross-polarization coefficient in reflection mode of more than 0.8 between 0.66-2.43 THz. In point of fact, polarization manipulation in transmission mode has evoked substantial attention because of their abundant usage in modern commercial applications. For highly practical transmissive converter, different types of structures have been suggested. For example, one kind of polarizer in transmission modes is based on graphene metasurface [25-27] which are switchable by electrically controlling the Fermi energy of the graphene sheet. Because of biasing devices that consist in their structure, these polarizers will have bulky configuration and high losses. Liu et al. [28] proposed a single-layer MS structure as a cross-linear polarizer that its functionality is based on the contribution of localized surface plasmon (LSP) and surface plasmon polaritons (SPPs). Although it appears that the fabrication process of multi-layer structures is more complicated than single layer ones, their intensity of the transmission wave is several times higher than the intensity of the transmission wave in single layer structures. In fact, single-layer meta-surface structures have limited interaction with the EM wave and subsequently will have lower efficiency that is essential for practical applications.

In this paper, we present a new design of a broadband and high-efficiency double-layer linear-cross polarizer based on periodic S-shaped resonators that placed on the bottom of the dielectric substrate. In addition, we proposed another cross-polarization converter operating in transmission mode that composed of an array of wire grid as a bottom metallic layer. Both of these polarizers are consisting of asymmetric split-ring resonators as a top metasurface layer. These polarizers can mainly convert the polarization of a linearly polarized (LP) wave into its orthogonal direction in transmission mode for the sub-THz region. The simulation results of the first structure show that the polarization conversion ratio (PCR) is over 85% in the frequency range from 86 to 139 GHz with a RBW of 46.4% and high cross-polarization transmission of 0.87 at 100GHz is achieved. The underlying reason for choosing a second structure, wire-grids structure, is its extremely low co-polarized transmission amplitude which is less than 0.03 and represents exceedingly a good polarization transformer in cross-direction. Furthermore, this structure has polarization conversion ration more than 95% in frequency from 50 to 200GHz, has a wide relative bandwidth. Moreover, polarization conversion functionality and the RBW of these polarizers remain nearly constant while the angle of the incident wave changes from $0°$ to $50°$. In comparison with previous polarizers[18], We can claim that our designed polarizers have several advantages of wider bandwidth, higher cross-transmission amplitude, ultrathin thickness of the structure, and also hav simple geometry. These proposed structures can find application in mm-wave and THz sensing and imaging systems which benefit from polarimetric measurements.

## 2. DESIGN OF CROSS-POLARIZER METASURFACE STRUCTURE

In this section, first, we will present a new design of a broadband and high-efficiency double-layer linear-cross polarizers based on periodic S-shaped resonators that placed on the bottom side of the dielectric substrate. Then, we proposed another cross-polarization converter operating in transmission mode that composed of an array of wire grid as a bottom metallic layer. Both of these polarizers are consisting of analyzer, asymmetric split-ring resonators, as a top metasurface layer. These polarizers can mainly convert the polarization of a linearly polarized wave into its cross-direction in transmission mode.

### 2.1. CROSS-POLARIZER WITH S-SHPAPE RESONATORS

Figure 1 (a), (b) illustrate the rectangular diagram of a single cell of the proposed periodic metasurface converter that is made up of ASRR and pairs of SRs deposed on two sides of the dielectric substrate with the thickness of $t\_s$. The periodicity of the structure along the *x* and *y* directions are $Px$ and $Py$. Asymmetry in the top metasurface layer (ASRR) is produced by two gaps with a dimension of *g* which are getting far apart from each other along the *x*-direction. In order to achieve the expected transmission amplitude, the thickness of the dielectric layer has been optimized and Rogers RT4-5880 with a standard thickness of $t\_s$=0.254mm, relative permittivity of 2.2, and loss tangent 0.0009 has been selected. In addition, the conductive layers on the top and bottom sides of the substrate are copper films with a thickness of 0.018mm. After carrying out the optimization process which will be shown later, the dimensional parameters of the polarizer are as follows: $Px$= 532 µm, $Py$ = 822 µm, *a* = 721.4 µm, *b* = 72.6 µm, *g*= 50 µm, *w1* = 60 µm, *w2* = 196.8 µm, $L_1$ =196 µm, *d* = 14 µm, *f*= 83 µm, *w3* = 68.4 µm, *h* = 143.3 µm, *p*=149.5µm, *q*=286.3µm, *r*=310.4µm, *s*=242µm, *k*=28.9µm, *m*=13.9µm.

### 2.2. CROSS-POLARIZER WITH WIRE-GRIDS STRUCTURE

In this part, we present another double-layer metasurface cross-polarizer which is composed of ASRR and metallic grating on two sides of the dielectric layer. The bottom side of the structure is shown in Fig.1 (c). In order to obtain wider bandwidth and lower co-polarization coefficient, and the best polarization conversion effect comparing to the previous polarizer, we substitute the second S-shape layer with a wire grating structure while the top layer as we have shown in Fig.1 (a), the characteristic of the dielectric layer, and the other parameters have remained unchanged. The wire-grid which is along the *x*-axes is transparent to the *y*-polarized wave while it will block the *x*-polarized wave. The all geometric parameter of the metallic copper is given by *w4*=289µm, $P_2$=113.6µm, and *c*=68.2 µm.

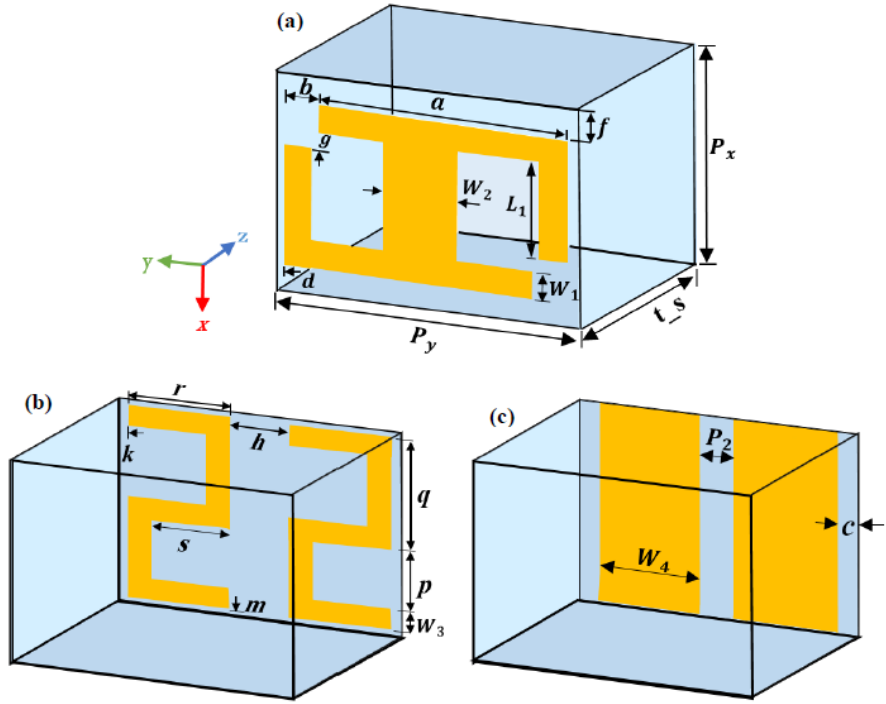

Fig. 1.Unit-Cell of proposed MS polarizers (a) ASRR side, the top layer of both polarizers (b) double SRs side (c) wire-grid side of second polarizer.

## 3. SIMULATION RESULTS AND DISCUSION

In order to elucidate the innovative functionalities underlying the suggested polarizers, relevant analysis of the unit cell has been performed by using the finite element method (FEM) in frequency domain solver of CST Microwave Studio. In our simulation, unit cell boundary conditions are considered along the *x* and *y* directions to characterize the periodic structure, and open (open-add space) along the *z*-axes to draw out scattering parameters.

Each layer in our proposed polarizers will have a significant effect on the working mechanism of the whole structure. As shown in Fig.2 (a), ASRR is an analyzer which is asymmetric with respect to the *x*-axis. Therefore, a portion of *x*-polarized incident wave converts to the *y*-polarization due to the induced current that forms electric dipole along the *y*-axes [30]. Subsequently, the simulation results of double SRs and wire-grid at Fig.2 (b) and Fig.2 (c) reveal that these structures are *y*-polarization selector. As shown in Fig. 2(b) and Fig.2 (c) below, the *y*-polarized incident wave can perfectly pass through these components while the *x*-polarized wave will be totally reflected. As a matter of fact, SRs are the evolving structure of wire-grids that are along *x*-axes [۳۱].

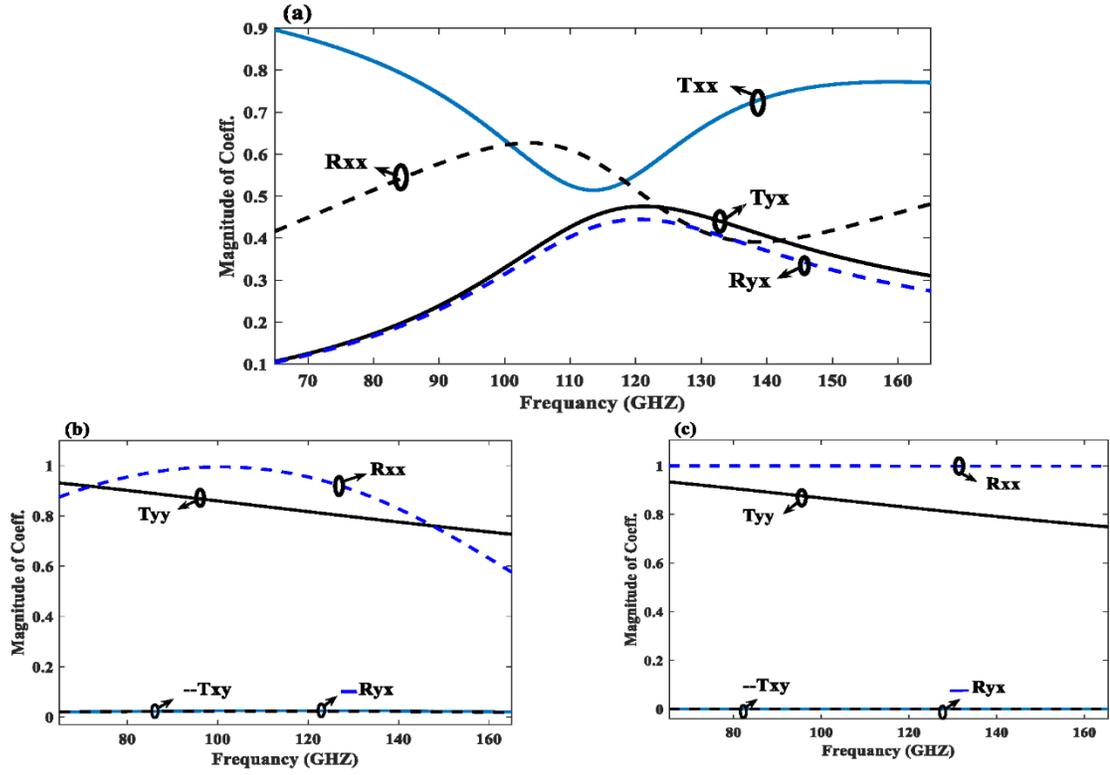

Fig. 2. (a) Simulated transmitted and reflection of (a) ASRR (b) double SRs and (c) wire-grid layer.

To validate these interpretations, the transmission coefficients and PCR which defines as $T_{yx}^2/(T_{yx}^2+T_{xx}^2)$ of the proposed structures are depicted in Fig. 3. Fig.3 (a) shows that the presented converter with S-shaped resonators is able to perfectly transform the polarization of a linearly *x*-polarized wave to its cross-transmitted wave in almost a wide frequency range and flat top curve. It can be observed that the cross-polarized transmission coefficient $(T_{yx})$ reaches 0.87 at 100GHz. In contrast, the co-polarized transmission coefficient $(T_{xx})$ is 0.06 at the same frequency which shows a complete polarization conversion ratio of 99.5%. The simulated PCR is shown in Fig.3 (a), too. It illustrates that, the proposed MS structure has a broad frequency band in which the PCR exceeds 85% from 86 to 138.9 GHz with a relative bandwidth of nearly 46.4%. In addition, from Fig.3 (b), it can be found that the co-polarized transmission amplitude of polarizer with wire-grid structure is extremely low and does not be more than 0.03 over the entire frequency ranges, however, the cross-polarized transmission amplitude is more than 0.8 from 92 to 140 GHz and it reaches 0.86 at a frequency of 100GHz. In addition, this figure indicates another remarkable point that the PCR of the *x*-polarized incident wave is above 95% over the entire-frequency and it reaches approximately near unity from 80 to180 GHz which means that this polarizer converts the *x*-polarized incident wave thoroughly to the *y*-polarized wave with great efficiency. We deliberately can choose SRs structure instead of the wire-grids, since SRs resonators will allow us to have denser ASRR array which will cause greater polarization convertibility. However, we should consider that the extremely low co-transmission of the wire-grid structure will cause pure cross polarized transmission wave in out-put of the structure.

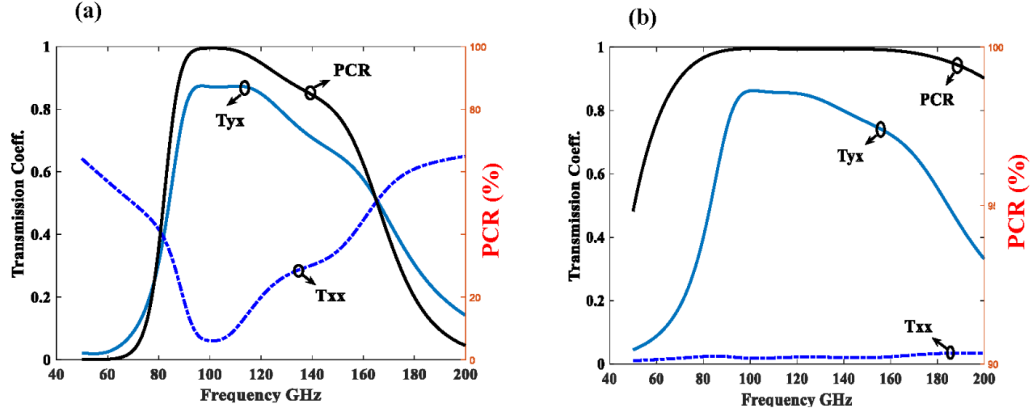

Fig. 3. Simulated result of the transmission coefficient and PCR for an incident wave that polarized along the *x*-axis for (a) polarizer with S-shaped resonators and (b) polarizer with wire-grid structure.

The total transmission for *x*- and *y*- polarization are defined as below [32]:

*x*-polarized incident wave: $T_x = |T_{xx}|^2 + |T_{yx}|^2$ (4)

*y*-polarized incident wave: $T_y = |T_{yy}|^2 + |T_{xy}|^2$ (5)

Fig .4(a) displays the total transmission of the *x*- and *y*- polarized incident wave for the polarizer which is consisted of S-shape resonators. It is obvious that the total transmission of the *x*-polarized wave is above 0.7 in the frequency range of 91.6 -127.4 GHz and the total transmission of *y*-polarized is below 0.25. This result indicates that; this polarizer transmits *x*-polarized waves well. In contrast, it doesn't permit the *y*-polarized wave transmits through the structure thoroughly. In fact, most of the *y*-polarized incident wave will be reflected from the structure.

Fig .4(b) shows the PCR of the structure under the *x*- and *y*-polarized incident wave. It is clear that the PCR of the *x*-polarized incident wave is above 85% from 86-139 GHz. However, the PCR of the *y*-polarization incident wave is under 20% at this range of frequency. It shows that, for the *y*-polarized incident wave, polarization conversion ratio reaches minimum value of zero at 100GHz and a maximum value of 20% at 85GHz. These results indicate that, this polarizer is a good polarization transformer under an *x*-polarized incident wave and broad frequency band. Approximately, the same results will be obtained for the second polarizer and we consider not show those results in order to eliminate replication.

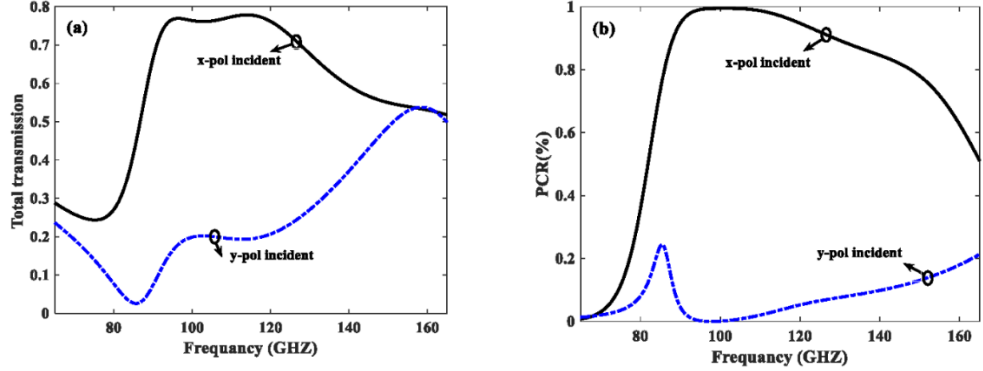

Fig. 4. (a) Total transmission and (b) PCR for *x*- and *y*-polarized incident wave

## 4. MORE PARAMETER STUDY AND SENSIVITY ANALYSIS

Some geometric parameters have a noteworthy impact on the performance of the proposed polarizers. In this section, first, we will optimize the performance of the structures by sweeping the value of some parameters like the width of $w_1$ and $w_4$, and the gap distance (*g*) as are shown in Fig. 1, and the effect of the number of the S-shaped resonators on the performance of the first proposed polarizer. Then, in order to investigate the sensitivity of the metasurface structure to the fabrication and implementation tolerances, we carry out numerical simulations by changing some other parameters like the thickness of the substrate layer ($t\_s$), thickness of copper layers ($t\_c$), the value of relative permittivity, the value of the incident angle $\theta_{in}$, and the influence of misalignment between the two layers of the structure. Numerical calculation with CST software will be carried out in order to obtain the best parameters for obtaining high conversion efficiency.

We have investigated the influence of the $w_1$ on the transmission coefficients in Fig.5 (a). As we can see, the operation frequency is shifted to a lower frequency when the width of $w_1$ increases. Subsequently, the structure has a cross-transmission coefficient above 0.8 in a wider frequency band for $w_1$=60µm. Through symmetry breaking of the ASRR array by displacing the two gaps from each other, a strong electric component of light which is perpendicular to the incident one will be achieved. To optimize adequate polarization conversion, we simulate the unit cell of the structure with different values of *g*. Fig. 5. (b) shows the variation of transmitted coefficients when the two gaps are varied from 15 to 85 µm while the other parameters are kept constant. It is obvious that the best optimum value for the gap length is 50µm in which maximum cross-transmission amplitude is achieved at a wide frequency range. Besides, Fig. 5(c) shows that, the underlying advantages of double S-shaped resonators to the single S-shaped resonator is its greater cross-transmission coefficient and wider bandwidth. In addition, Fig.5 (d) shows the optimization process of the wire-grid. It is crystal clear that, when $w_4$ increases from 110 to 289 µm the amplitude of the cross-transmission coefficient will gradually increase and it reaches to its maximum amount at 100GHz. Moreover, the cross-transmission coefficient is above 0.8 in wider frequency range when $w_4$ is 289 µm.

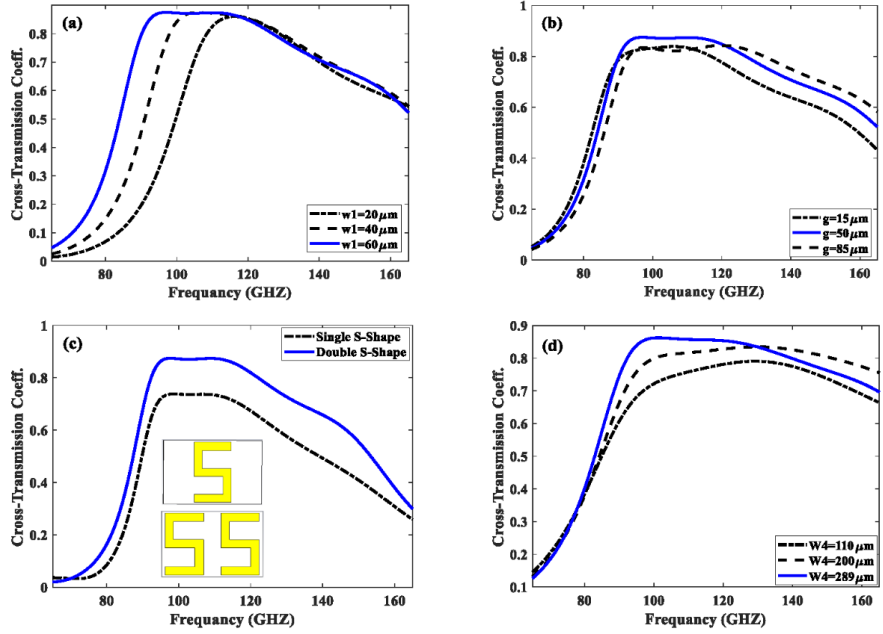

Fig. 5. The simulated Cross-Polarized transmitted amplitude for different value of (a) $w_1$ from 20 to 60μm (b) asymmetric gaps from 15 to 85μm (c) different number of the S-shaped resonators and (d) $w_4$ from 110 to 289 μm.

Fig.6. shows the influence of the parameters which have tolerance during the fabrication and implementation processes on the performance of the proposed polarizer with S-shape resonators. Fig. 6(a) shows the simulated results of the different standard thickness of substrate layer ($t\_s$) which are available and common, under the normal incidence of the *x*-polarized wave. It clearly indicates that most of an incident wave will be converted to the cross-polarized transmitted wave when $t\_s$ is 254, especially at 100GHz. additionally, the cross-transmission coefficient remains flat and above 0.8 under a broader frequency range which shows a remarkable asymmetric transmission effect for a linearly polarized wave. With the increase of the *t*, the cross-polarization coefficient descends from 0.87 to 0.5 at frequency of 100GHz. Furthermore, it is important to investigate the performance of the proposed structure for different copper thickness ($t\_c$). The simulated cross-transmission is given in Fig.6 (b). Simulated result reveals that by increasing the thickness of copper layers from 9 to 35μm, this metal claddings are available based on dielectric thickness, the cross-polarized transmission coefficient and the RBW will increase gradually. Due to the restrictions of fabrication equipment, we select a copper layer with a thickness of 18μm. In order to evaluate the influence of the material of the substrate on the transmission effect, we simulate the variation of cross-coefficient with the frequency by different values of relative permittivity that has changed with the tolerance of $\pm 10$ in Fig.6 (c). Figure below shows that the operation frequency is shifted to the lower frequencies when the $\varepsilon_r$ increases and also causes greater cross-transmission amplitude in 100GHz.

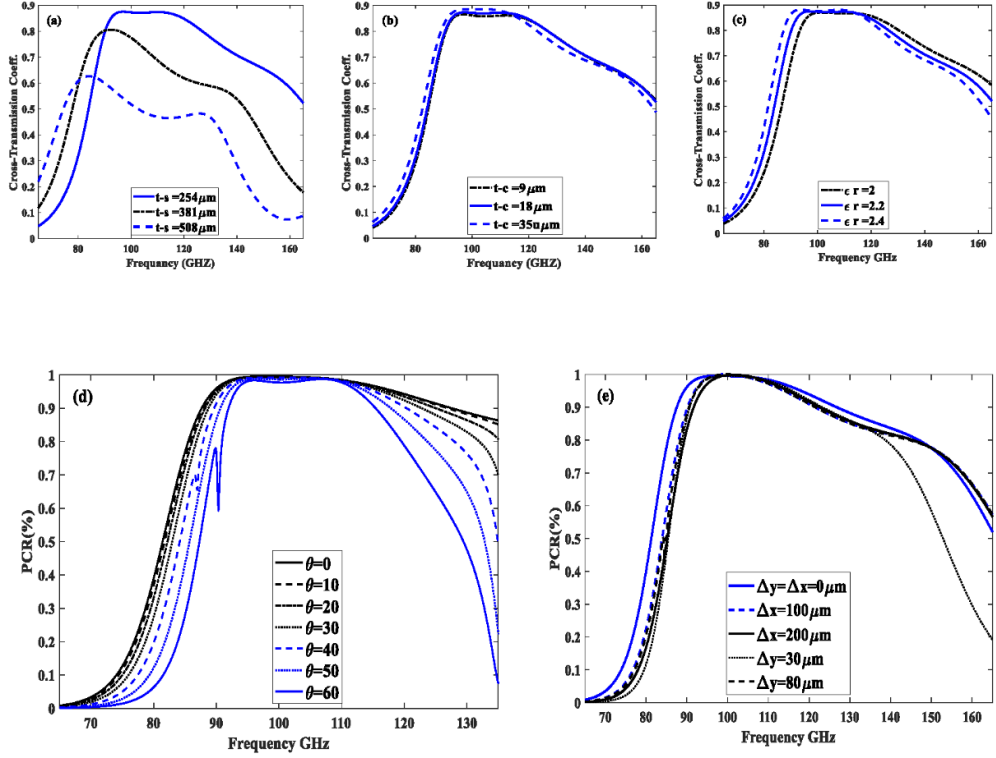

Fig. 6. The simulated Cross-Polarized transmitted amplitude for different value of (a) thickness of dielectric layer (b) different thickness of copper layers and (c) different value of relative permittivity. Simulated PCR of the proposed cross-polarizer (d) at oblique incident EM wave with different value of $\theta$ from $0°$ to $60°$ (e) and different misalignment between ASRR a SRs.

Furthermore, the PCR of this proposed polarizer as a function of frequency has been simulated under different values of incident angle $\theta_{in}$ inoder to demonstrate the incident angle insensitivity of the proposed structure. The simulated result, shown in the figure. 6(d), indicates that the functionality and the RBW of the polarizer maintain nearly constant while the angle of the incident wave changes from $0°$ to $50°$. It shows that our designed polarizer performs excellently under a wide range of incident angles. However, the bandwidth and the value of PCR are gradually declined with increasing the incident angle larger than $50°$. Herein, we have examined the influence of misalignment between the two layers of the structure. Each pair of SRs were displaced in the x-direction with $\Delta x$ or in y-direction with $\Delta y$, with respect to each ASRR. Fig.6 (e) reveals the simulated results of misalignment. It indicates that the performance of the misalignment structures is nearly at good consistence with well-aligned one and displacement will not have an influence on the functionality of the whole structure at the frequency range around the 100GHz and it also facilitates the fabrication process. While, it should be noticed that the rotational orientation of the layers need to be aligned because both of them are sensitive to the polarization of an incident wave. We have investigated the effect of these parameters on the second polarizer. In fact, the same results as the above results have been obtained so, we do not repeat them for non-repetition.

## 3. FABRICATION AND MEASUREMENT

Finally, to demonstrate an experimental validation of these designed polarizers, a prototype of each polarizer which covers an area of about 10×10 mm$^2$ is fabricated by lithography technology. These polarizers are fabricated on a double-sided raw PCB with 18 µm Copper on each side and 2٠٤ µm RT4-5880 dielectric in between. A thin layer of positive photoresist is spin-coated on one of the Copper films. The desired pattern is transferred to the photoresist by standard optical lithography. Then, the sample is etched in 20% FeCl3+ 10% HCl at 55°C while the Copper film on the other side is protected with an etch-resistant ink. The whole process should be repeated on the second side while the first side is protected in the etching step. Photos of the two sides of both fabricated polarizers can be seen in Fig. 7.

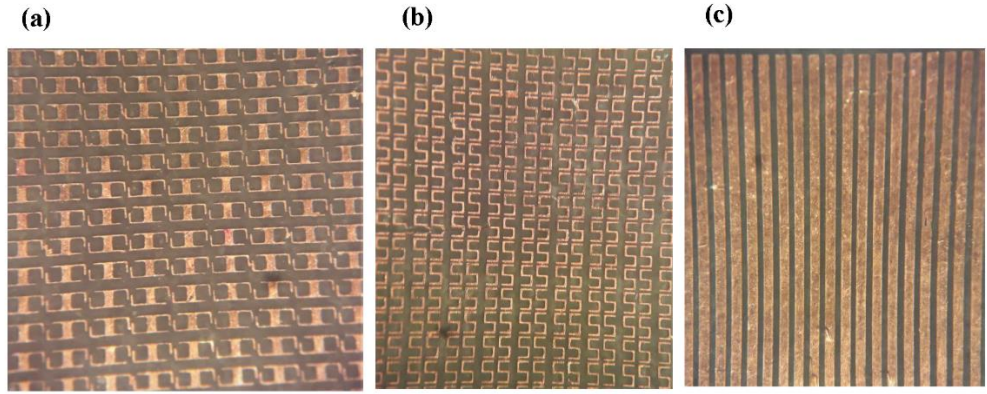

Fig. 7. Images of fabricated sample (a) viewed from ASRR side of both polarizer (top side) (b) viewed from Double SRs side for the polarizer consist of S-Shaped resonator (bottom side) (c) viewed from Wire-Grid side for the polarizer consist of wire-grid (bottom side).

The transmission coefficients of the prototype metasurfaces are measured by using a continuous-wave (CW) THz source and detector which is schematically shown in Fig.8. The femtosecond laser diode with the emission wavelength of 1550nm and bandwidth of 40nm drives a photoconductive antenna. The antenna structure is dipole shape which emits THz continuous-wave containing frequency components from 50 GHz to 2.7 THz. The THz beam emitted by the photoconductive antenna is guided via four off-axis parabolic mirrors PM1-PM4. Mirrors PM1 and PM3 are used for beam collimation, while PM2 and PM4 are for beam focusing. The prototype polarizer is located between the mirrors PM3 and PM2. Eventually, the transmitted wave from the polarizer is focused on the coherent detector antenna. For measuring the co-transmission coefficient, the polarization of the emitter and detector are in the same direction. In contrast, the detector is rotated 90° to get the cross-polarization coefficient because it can receive only the $x$-polarization. In addition, the transmission was also measured with a vacant window without the presentation of sample, and cross- and co-polarization transmission were both normalized by it.

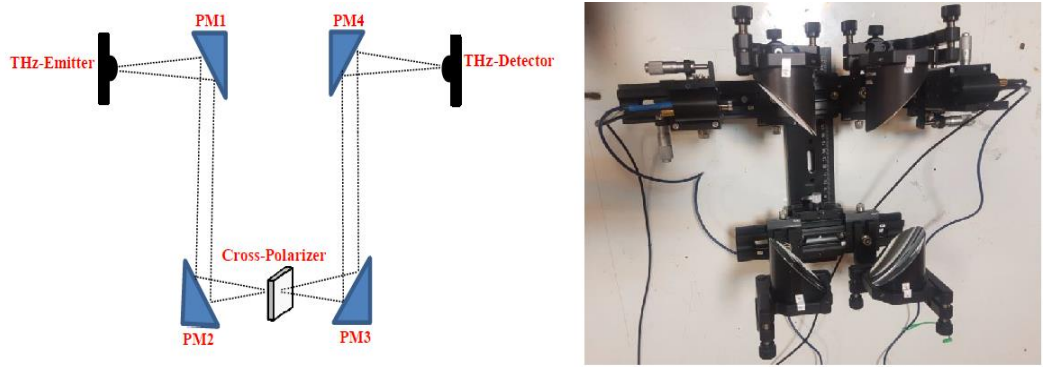

Fig. 8. Schematic diagram of measurement setup based on THz-CW system consist of a photoconductive emitter and detector along with for off-axis parabolic mirrors for collimation and focusing.

Fig.9 shows both the measurements of the fabricated samples and the simulation results of our proposed polarizers. It indicates that the experimental results are almost in good agreement with simulation results which shows the validity of our proposed polarizers. Both simulated and experimental results reveal that these proposed metasurfaces can convert a linearly $x$-polarized wave to its orthogonal direction in a wideband sub-THz region. We should consider that the frequency shifts and small dissimilarity are reasonable due to errors that could occur in measurement implementation and fabrication process. Besides, the size of prototype samples are very limited compared to infinite structures that are considered in simulation and pure incident plane wave cannot be achieve in measurement setup.

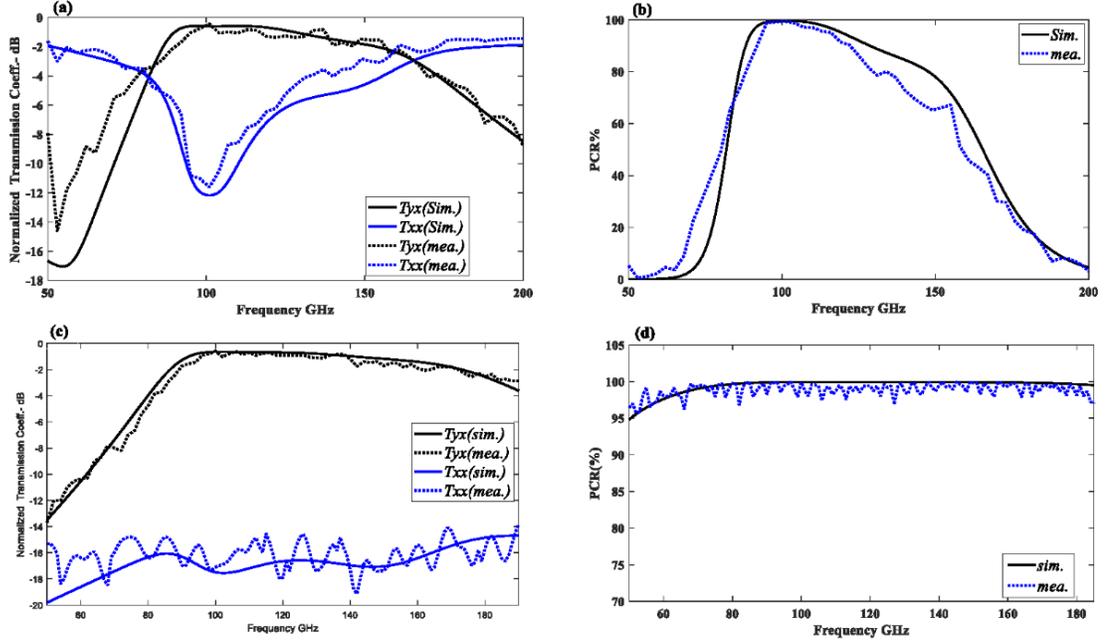

Fig. 9. Measurement and simulation results of (a) co- and cross-transmission coeff. (b) PCR for polarizer with S-Shape resonator (c) cross-transmission coeff. (b) PCR for polarizer with wire-grid versus frequency.

## 4. CONCLUSION

In conclusion, in this work, we have proposed and fabricated different broadband and high-efficiency metasurface-based cross-polarization converters for the sub-THz region based on ASRR as a top metasurface layer. The cross-polarizer with SRs resonators produce high-asymmetric transmission for a linearly polarized incident wave in transmission mode. Simulation and experiment show that a RBW of 46.4% from 86 to 138.9 GHz in which the PCR of the *x*-polarized incident wave is more than 85% is obtained. The superiority of this proposed structure is its filtering property and high conversion efficiency at its resonant frequency of 100 GHz which its cross-polarized transmission coefficient reaches 0.87. Subsequently, we present other wideband linear polarizer which has near-perfect and unity PCR for the *x*-polarized incident wave in most part of the frequency region. Besides, the co-polarized transmission amplitude is extremely low and does not be more than 0.03 over the whole-frequency ranges. Furthermore, these polarizers keep high-efficient polarization conversion over a wide frequency band when the incident angle of *x*-polarized EM wave alters from $0°$ to $50°$. In addition, the tolerance of the structure to the misalignment between the two layers makes the fabrication process easier. Finally, the experimental results of these polarizers show good accordance with the simulation ones. Although these cross-polarizers works for the sub-THz region, it is also realizable to microwave and even optical frequencies by scaling its geometric dimensions. These kind of polarizers can have potential applications in the sub-THz imaging system, polarimetry and etc.

Table.1 compares significant characteristics of our two novel polarizers with recently published metasurface based ones. Obviously, it can be concluded that our proposed polarizers have a wide frequency band in which their PCR are perfect in the sub-THz region and they have a high magnitude of asymmetric transmission coefficient. Our designed polarizers have thin thickness, and they also have a simple geometry. Having these significant advantages simultaneously are their superiority in comparison with other polarizers which are reported below.

Table 1. Comparison between our proposed polarizer with other reported polarizers

| Ref. | Device Structure | PCR Peaks% | Cross-Pol Peaks | PCR | RBW(CF) | Max In Angle |
|---|---|---|---|---|---|---|
| [18] | Double Layer | 99.9% | 0.71 | > 0.85<br>0.94 – 1.13THz | 18.35%<br>(1.035THz) | Not-Mentioned |
| [21] | Tri-Layer | Not-Mentioned | 0.8 | > 50<br>0.52 – 1.82THz | 111%<br>(1.17THz) | 45º |
| [28] | Single Layer | 95% | ~ 0.3 | > 0.9<br>0.91 – 1.45THz | 45.7%<br>(1.18THz) | Not-Mentioned |
| **Polarizer with S-shape** | Double Layer | 99.5% | 0.87 | > 0.85<br>86 – 138.9GHz | 46.4%<br>(112GHz) | 50º |
| **Polarizer with Wire-Grid** | Double Layer | 99.9% | 0.86 | > 0.95<br>50-200GHz | 120%<br>(125GHz) | 50º |

CF= Central Frequency
BW= Bandwidth
RBW=Relative Bandwidth